\documentclass[preprintnumbers,article,floatfix,10pt,prd,onecolumn,superscriptaddress,nofootinbib]{revtex4-2}
\usepackage{bm}
\usepackage{amsfonts}
\usepackage{tikz}
\usepackage{latexsym}
\usepackage[latin1]{inputenc}
\usepackage{graphicx}
\usepackage{amsmath}
\usepackage{palatino}
\usepackage{mathpazo}
\usepackage{textcomp}
\usepackage{float}
\usepackage{booktabs}
\usepackage{dcolumn}
\usepackage{ragged2e}
\usepackage{hyperref}
\usepackage{xcolor}
\usepackage{epsfig}
\usepackage[caption=false]{subfig}
\usepackage{commath}
\usepackage{multirow}

\hypersetup{colorlinks,citecolor=blue}
\hypersetup{colorlinks=true,linkcolor=red,filecolor=magenta,urlcolor=blue}
\newcommand{\be}{\begin{equation}}
\newcommand{\ee}{\end{equation}}
\newcommand{\bea}{\begin{eqnarray}}
\newcommand{\eea}{\end{eqnarray}}
\newcommand{\eeas}{\end{eqnarray*}}
\newcommand{\beas}{\begin{eqnarray*}}

\allowdisplaybreaks[1]

\begin{document}

\title{Exploring the Dark Age: Star and Galaxy formation in the Early
Universe}
\author{K. El Bourakadi\,\href{https://orcid.org/0000-0002-2199-9613}{%
\includegraphics[width=8pt]{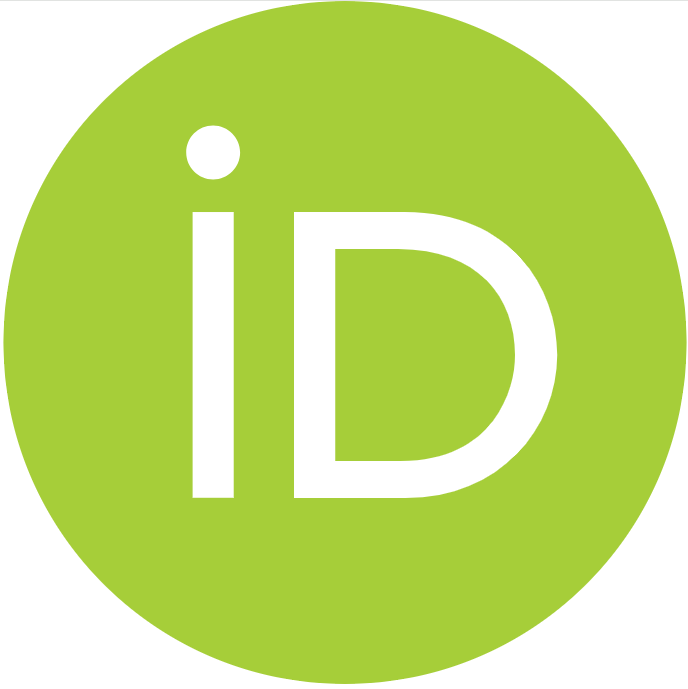}}}
\email{k.elbourakadi@yahoo.com}
\affiliation{Departamento de F\'isica, Facultad de Ciencias, Universidad de
Tarapac\'a, Casilla 7-D, Arica, Chile.} 
\affiliation{Subatomic Research and
Applications Team, Faculty of Science Ben M'sik, Casablanca Hassan II
University, Morocco.}
\author{G. Otalora\,\href{https://orcid.org/0000-0001-6753-0565}{%
\includegraphics[width=8pt]{orcid.png}}}
\email{giovanni.otalora@academicos.uta.cl}
\date{\today }
\affiliation{Departamento de F\'isica, Facultad de Ciencias, Universidad de
Tarapac\'a, Casilla 7-D, Arica, Chile.}

\begin{abstract}
The Cosmic Dark Ages mark a pivotal era of the universe's evolution,
transitioning from a neutral, opaque medium to the emergence of the first
stars and galaxies that initiated cosmic reionization. This study examines
the thermodynamics of the intergalactic medium (IGM), molecular hydrogen
cooling, and gravitational collapse that led to structure formation. Key
emission lines, such as Lyman-alpha (Ly$\alpha $) and [C II] 158 $\mu m$,
are analyzed as tracers of star formation, metallicity, and IGM conditions.
Simulations highlight Ly$\alpha $ scattering profiles and [C II] emission as
critical diagnostics of early galaxy evolution. The findings provide a
theoretical framework to interpret high-redshift observations, advancing our
understanding of the universe's transition from darkness to illumination.
\end{abstract}

\maketitle

\section{Introduction}

\label{sec1}

Observations from cutting-edge telescopes across $\gamma $-ray to radio
wavelengths, astronomers are now uncovering galaxies, Active Galactic Nuclei
(AGN), which trace back to roughly 500--940 million years following the Big
Bang \cite{I1}. This period marks the era of "cosmic reionization," during
which radiation from the first galaxies and black holes reionized the
neutral InterGalactic Medium (IGM) that dominated the universe after
recombination. Observational evidence, such as data from the cosmic
microwave background \cite{I2}, the Gunn--Peterson trough in the spectra of
quasars at $z>6$ \cite{I3}, the characteristics of Ly$\alpha $ emission in
galaxies at similar redshifts \cite{I4}, and constraints on the 21 cm signal
from the neutral IGM \cite{I4,I6}, has increasingly refined the timeline of
reionization. It is now evident that the transformation of the IGM from
predominantly neutral to highly ionized occurred primarily between redshifts 
$z\sim 6$ and $10$, driven by the emergence of early galaxies \cite{I7,I8,I9}%
. The next decade promises significant advancements in understanding this
critical phase of cosmic evolution, thanks to the James Webb Space Telescope
(JWST), upcoming 30-meter-class ground-based telescopes for optical and
near-infrared studies, and the expanded capabilities of the Atacama Large
Millimeter Array (ALMA) across its full frequency range. These cutting-edge
tools are set to provide unprecedented insights into this pivotal era.

The universe at $z\sim 15$ represents a frontier in our current
understanding of cosmic evolution. Few theoretical investigations have
explored the Star Formation Rate (SFR) density at these extreme redshifts,
often in the context of early reionization \cite{I10,I11,I12,I13,I14}.
Unlike galaxy formation at lower redshifts, early structures are thought to
have significantly lower dynamical masses, making them more vulnerable to
supernova feedback, which could inhibit or completely halt star formation by
ejecting or heating their gas. Additionally, radiative feedback from nearby
sources producing photoionizing radiation increases the Jeans length in the
intergalactic medium, preventing the formation of the smallest galaxies with
circular velocities below $\sim 50km/s$ \cite{I15}. These processes are
expected to have an increasingly significant impact as we probe into higher
redshifts. The cosmic infrared background (CIRB) provides valuable insights
into the star formation history of the universe \cite{I16}. This background
radiation is a blend of emissions from all extragalactic sources. In the
near-infrared (NIR) range, the CIRB primarily originates from stellar light,
while emissions at mid- and far-infrared wavelengths are predominantly
driven by the thermal radiation from cosmic dust. Several researchers have
measured the near-infrared background (NIRB) in the J, K, and L bands using
data from the Diffuse Infrared Background Experiment (DIRBE) aboard the
Cosmic Background Explorer (COBE) satellite. More recently, Ref. (\cite{I17}%
) provided precise measurements of the NIRB across the 1.4 to 4 $\mu $m
range using data from the near-infrared spectrometer (NIRS). A significant
challenge in these measurements arises from uncertainties related to
subtracting sunlight scattered by interplanetary dust (IPD). Despite this
issue, some studies suggest that the observed NIRB cannot be fully explained
by contributions from ordinary galaxies \cite{I18}.

A previous study \cite{I19} proposed that the earliest generation of stars,
known as Population III (Pop III) stars, could contribute to the
near-infrared cosmic background. Later, Ref.(\cite{I20}) expanded on this
idea, suggesting that a substantial portion of the unexplained near-infrared
background might originate from Pop III stars. They explored a scenario with
a highly top-heavy initial mass function (IMF), where stars are assumed to
have masses around 300 $M_{\odot }$, as indicated by recent
three-dimensional simulations \cite{I21}.

The exploration of the interstellar medium (ISM) in early-universe galaxies
has entered an exciting phase, thanks to the remarkable capabilities of the
ALMA. One of the most significant tools in this study is the 158 $\mu m$
emission line associated with the $^{2}P_{3/2}\rightarrow ^{2}P_{1/2}$
fine-structure transition of ionized carbon ([C II]). This line, as the
primary cooling mechanism for neutral diffuse ISM \cite{I22}, is the
brightest emission feature in the far-infrared spectrum \cite{I23}. Beyond
the neutral diffuse gas, the [C II] line can also arise from high-density
photodissociation regions (PDRs) and diffuse ionized gas, where free
electrons drive the emission through collisions. While disentangling the
contributions from these various gas phases can be challenging, the [C II]
line remains a powerful diagnostic for studying the ISM in galaxies from the
Epoch of Reionization (EoR; $z=6-7$; e.g., \cite{I24}). Prior to ALMA,
detections of the [C II] line at $z>4$ were limited to galaxies with extreme
star formation rates \cite{I24,I25,I26} or those hosting active galactic
nuclei \cite{I27,I28,I29,I30}.

In this work we explore the transformative epoch of the Cosmic Dark Ages,
detailing the universe's progression from a neutral, lightless state to the
emergence of the first stars and galaxies, driving cosmic reionization. Key
topics include the role of Population III and II stars in shaping early
intergalactic medium conditions, the influence of molecular hydrogen cooling
on gravitational collapse, and the critical role of emission lines like Ly$%
\alpha $ and [C II] in tracing star formation, metallicity, and ISM
properties. By combining theoretical models and computational simulations,
the paper provides insights into the emission and spatial distribution of
early stellar populations, the evolution of cosmic infrared background
features, and the comparative analysis of Ly$\alpha $ and [C II] emission in
high-redshift galaxies. This paper is subdivided as follows: in the Sec.\ref%
{sec2}, we discuss the Dark Ages and the mechanisms of structure formation.
In the Sec.\ref{sec3} , we investigate Population II stars, their emission
properties, and spatial distribution. Sec.\ref{sec4} examines Ly$\alpha $
and [C II] emissions as tracers of galaxy evolution during the epoch of
reionization. Finally, we conclude in Sec.\ref{sec5}, summarizing the key
findings and their implications for understanding high-redshift galaxy
formation and cosmic evolution.

\section{The Cosmic Dark Ages}

\label{sec2}

\subsection{The Dark Ages \& Structure Formation}

The Dark Ages refer to a period in the early universe that follows
recombination and the universe was predominantly devoid of luminous sources.
This era was marked by a neutral hydrogen-dominated intergalactic medium,
with minimal radiation capable of ionizing atoms or illuminating the cosmic
landscape \cite{A1}. Despite the lack luminous sources, density fluctuations
seeded by quantum effects in the inflationary period grew through
gravitational attraction, leading to matter clumping and the eventual
formation of primordial structures \cite{A2}. As density variations in the
cosmic gas and dark matter persisted, these regions with slight
over-densities began the gravitational collapse, leading to galaxy formation 
\cite{A3}. This gravitational collapse mechanism would later lead to the
formation of the first stars (Population III stars) in localized dense
regions, an event prepare for the end of the Dark Ages and the start of
cosmic reionization. During the Dark Ages, the IGM was primarily composed of
neutral hydrogen atoms with a trace amount of helium and tiny fractions of
light elements formed during Big Bang nucleosynthesis \cite{A4}. The gas
temperature decreased as the universe expanded, cooling to a point where it
was no longer sufficient to sustain ionized particles. This neutral state
persisted throughout the Dark Ages, impacting the opacity of the medium and
allowing neutral hydrogen to dominate the cosmic landscape. As outlined by
Ref. (\cite{A1}), the cooling of gas through mechanisms such as hydrogen $%
H_{2}$ formation was essential to initiate future star formation and
enabling gas clouds to fragment and collapse.

Neutral hydrogen played a critical role in shaping the spectral features
during this epoch through its Lyman-alpha $\left( Ly\alpha \right) $
absorption, affecting any radiation passing through the IGM. As seen in Ref.
(\cite{A5}), the effective optical depth $\left( \tau _{IGM}\right) $ and
higher-order Lyman series transitions was significant which imposed a
blanket absorption on radiation across these wavelengths. Only wavelengths
redshifted beyond the Lyman limit passed without obstruction, marking the
absorption characteristic of the Dark Ages. The scattering and reemission of
photons through Lyman series contributed to faint diffuse emissions that
were simulated in Ref. (\cite{A6}), the results indicate that scattered $%
Ly\alpha $ photons developed a characteristic asymmetric emission profile
due to the IGM's Hubble expansion, creating a faint background glow that
permeated the Dark Ages. The background radiation intensity observed in the
cosmic infrared background (CIRB) is derived from the cumulative emissions
and absorptions throughout cosmic history. Peebles' formalism \cite{A5}
gives the specific intensity $I\left( \nu _{0},z_{0}\right) $ observed at a
frequency $\nu _{0}$ and redshift $z_{0}$, integrating over all redshifts $%
z>z_{0}$. For the Dark Ages, the emissivity $\epsilon \left( \nu ,z\right) $
primarily reflects the IGM's thermal properties and weak emissions from
non-ionizing processes:%
\begin{equation}
I\left( \nu _{0},z_{0}\right) =\frac{1}{4\pi }\int_{z_{0}}^{\infty }\epsilon
\left( \nu ,z\right) e^{-\tau _{IGM}\left( \nu _{0},z_{0},z\right) }\frac{dl%
}{dz}dz,
\end{equation}%
where $\frac{dl}{dz}=c\left[ H_{0}\left( 1+z\right) E(z)\right] ^{-1}$ is
the line element for proper distance, governed by cosmological expansion
parameters with $c$ is the speed of light and,

\begin{equation}
E(z)=\sqrt{\Omega _{M}(1+z)^{3}+\Omega _{\Lambda }+(1-\Omega _{M}-\Omega
_{\Lambda })(1+z)^{2}}.
\end{equation}%
is the dimensionless Hubble parameter. The above form demonstrates how
density fluctuations, Hubble expansion, and atomic transitions impact the
observable universe's spectral features. In this context, The effective
optical depth, denoted as $\tau _{IGM}$, through the IGM is defined by \cite%
{A7,A8},%
\begin{equation}
\tau _{IGM}\left( \nu _{0},z_{0},z\right) =\int_{z_{0}}^{z}dz^{\prime
}\int_{0}^{\infty }dN_{H_{I}}\varsigma \left( N_{H_{I}},z^{\prime }\right)
\left( 1-e^{-\tau }\right) ,
\end{equation}%
where $\varsigma \left( N_{H_{I}},z^{\prime }\right)
=d^{2}N/dN_{H_{I}}dz^{\prime }$\ represents the distribution of absorbers as
a function of redshift and neutral hydrogen column density, $N_{H_{I}}$. $%
\tau (\nu )$ denotes the optical depth of a single cloud for ionizing
radiation at frequency $\nu $.

\subsection{Population II Emission and Spatial Distribution}

The evolution and emission properties of Population II stars are critical
for understanding the cosmic timeline, especially during the early phases of
galaxy formation \cite{A9}. Population II stars, characterized by low
metallicity, exhibit distinct stellar and nebular emission signatures
influenced by their masses, metallicities, and evolutionary stages \cite{A10}%
. This section presents a mathematical formalism for modeling these
emissions and visualizes the spatial distribution of stars within a galaxy,
based on a computational simulation. Stellar emission arises from the
intrinsic radiative output of stars. For a star of mass $M$, metallicity $Z$%
, and age $t$, the effective temperature $T_{eff}$ and luminosity $L_{\ast }$
are modeled as\cite{A11,A12},%
\begin{eqnarray}
T_{eff} &=&6000\left( \frac{M}{1M_{\odot }}\right) ^{0.5}\left( 1-Z\right) ,
\\
L_{\ast } &=&M^{3.5}\exp \left( -\frac{t}{1Gyr}\right) .
\end{eqnarray}%
where the stellar emission is given by $E_{\ast }=T_{eff}\cdot $\ $L_{\ast }$
\cite{A13}. Nebular emission is driven by ionizing photons that interact
with the surrounding medium \cite{A14}. The ionizing photon rate $q_{H}$
scales with the stellar mass and metallicity as \cite{A15},%
\begin{equation}
q_{H}=10^{46}\left( \frac{M}{1M_{\odot }}\right) ^{1.5}\left( 1-Z\right) .
\end{equation}%
The resulting nebular emission, $E_{nebular}$\ is expressed as \cite{A16},%
\begin{equation}
E_{nebular}=c_{H_{\alpha }}\cdot \alpha _{B}\cdot q_{H}\cdot \frac{\exp
\left( -\frac{t}{1Gyr}\right) }{1-f_{esc}},
\end{equation}%
where $c_{H_{\alpha }}$ represents the energy output in the $H_{\alpha }$
spectral line, $\alpha _{B}$ is the recombination coefficient for case B,
which accounts for recombination events leading to photon emission in
excited states, and $f_{esc}$ denotes the fraction of ionizing photons that
escape the surrounding nebula \cite{A17}. The combined emission is modeled
as the sum of stellar and nebular components $E_{total}=E_{\ast
}+E_{nebular} $ \cite{A17}.

\begin{figure}[h]
\centering
\includegraphics[width=0.9\textwidth]{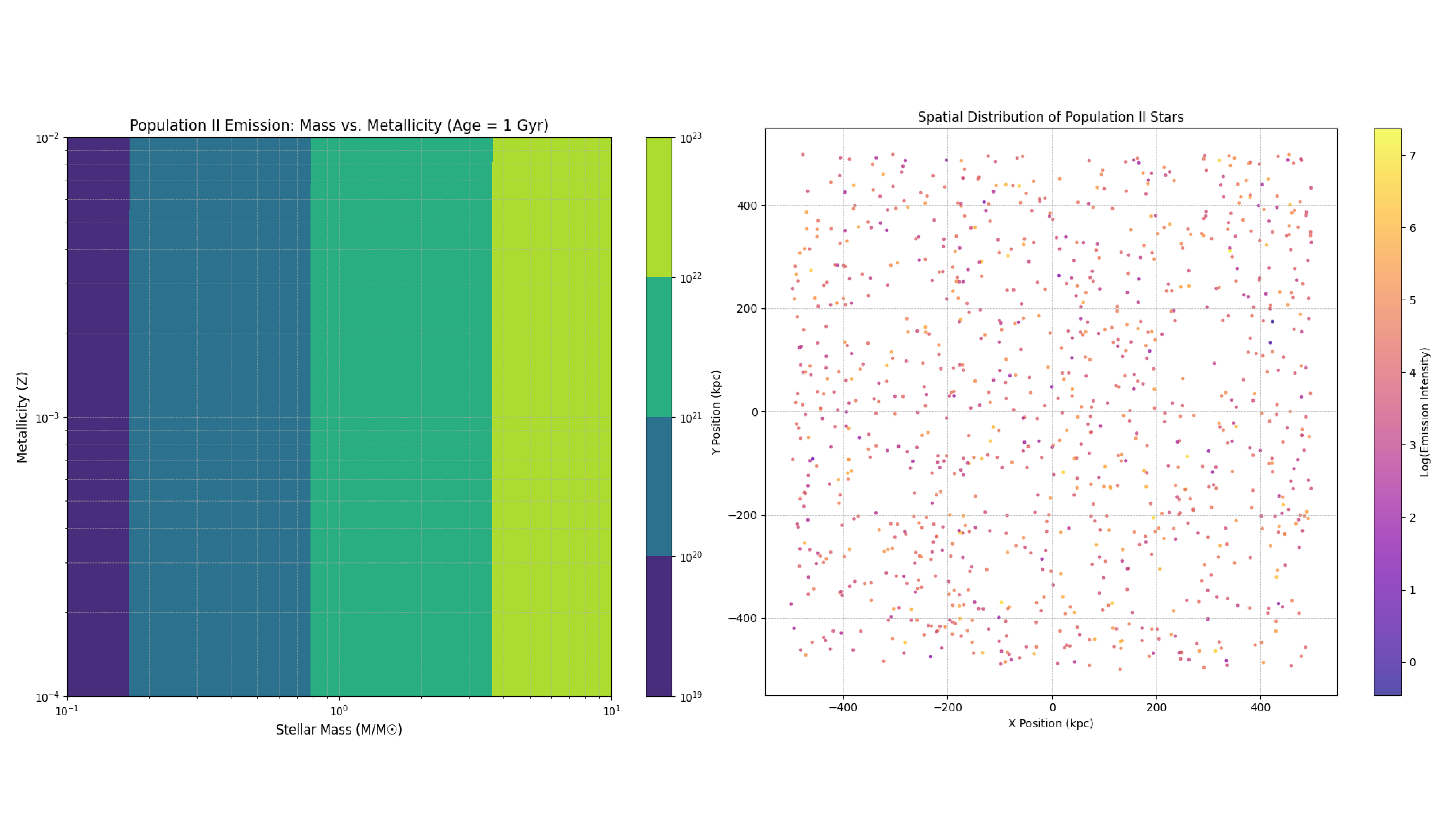} 
\caption{ Emission Intensity and Spatial Distribution of Population II Stars 
}
\label{fig:0a}
\end{figure}

Fig. \ref{fig:0a} illustrates the emission intensity of Population II stars
as a function of stellar mass ($M/M_{\odot }$) and metallicity ($Z$) at $%
1Gyr $. High-mass, low-metallicity stars dominate the emission due to their
high luminosity and ionizing photon production, driving stronger nebular
contributions. Emission decreases for lower-mass stars and at higher
metallicities due to reduced luminosity and effective temperature, resulting
in a diagonal trend of peak intensity. This pattern underscores the
significant role of massive, metal-poor stars in reionization and early
universe enrichment. The scatter plot depicts the spatial distribution and
emission intensity of Population II stars in a simulated $1kpc$ galaxy.
High-emission stars are sparse, reflecting the rarity of massive stars,
while low-emission stars dominate due to their abundance. The uniform
spatial distribution indicates an isotropic placement of stars, with mixed
contributions from short-lived, high-mass stars and long-lived, low-mass
stars. This highlights the diversity and radiative impact of Population II
stars on the light output of early galaxies.

The Dark Ages set the cosmic stage for all subsequent structure formation.
The growth of matter density fluctuations into gravitationally bound
structures was essential for the emergence of galaxies and clusters. This
period's analytical treatment of density perturbations, the thermodynamics
of the IGM, and the role of molecular hydrogen cooling are pivotal for
understanding the initiation of star formation. The gravitational collapse
criteria, minimum cooling thresholds, and subsequent fragmentation provide a
foundation for analyzing the transition into the era of first light and
cosmic reionization. This conceptual and analytical overview of the Dark
Ages offers insights into the universe's silent and foundational period,
paving the way for the birth of the first stars and cosmic illumination.

\subsection{Emission and Scattering in the Cosmic Infrared Background}

The Cosmic Infrared Background is shaped by the combined effects of Ly$%
\alpha $ photon scattering in the intergalactic medium and the emission from
first-generation Population III stars \cite{C1}. Ly$\alpha $ photons emitted
by these early galaxies undergo scattering interactions with the neutral
hydrogen in the IGM, diffusing to longer wavelengths due to the Hubble
expansion \cite{C2}. This results in an asymmetric Ly$\alpha $ emission line
with a scattering profile given by \cite{C3}:

\begin{equation}
\Phi (\nu ,z)=%
\begin{cases}
\nu _{\star }(z)\nu ^{-2}\exp \left( -\frac{\nu _{\star }}{\nu }\right) , & 
\nu >0, \\ 
0, & \nu \leq 0,%
\end{cases}%
\end{equation}%
where $\nu _{\star }(z)$, the characteristic frequency, depends on baryonic
density and cosmic expansion \cite{C3}, 
\begin{equation}
\nu _{\star }(z)=1.5\times 10^{12}\,\text{Hz}\,\left( \frac{\Omega _{b}h^{2}%
}{0.019}\right) \left( \frac{h}{0.7}\right) ^{-1}\frac{(1+z)^{3}}{E(z)},
\end{equation}%
Emission from Population III stellar clusters combines contributions from
stellar spectra, nebular emissions, and Ly$\alpha $ line emission. The total
emission spectrum is expressed as \cite{C4}, 
\begin{equation}
l_{\nu }(z)=\int_{M_{l}}^{M_{u}}F(\nu ,M,z)\phi (M)dM,
\end{equation}%
where $F(\nu ,M,z)$ includes stellar ($l_{\nu }^{\text{star}}$), nebular ($%
l_{\nu }^{\text{neb}}$), and Ly$\alpha $ line ($l_{\nu }^{\text{Ly}\alpha }$%
) contributions. Nebular emissions are described by \cite{C4}, 
\begin{equation}
l_{\nu }^{\text{neb}}=\gamma _{\text{tot}}\frac{\alpha _{B}}{1-f_{\text{esc}}%
}q(H),
\end{equation}%
with $\alpha _{B}$ as the Case B recombination coefficient, $f_{\text{esc}}$
the ionizing photon escape fraction, and $q(H)=Q(H)/M$ representing the
ionizing photon rate per unit stellar mass. The Ly$\alpha $ emission,
modulated by IGM scattering, is given by \cite{C5} 
\begin{equation}
l_{\nu }^{\text{Ly}\alpha }(z)=c_{\text{Ly}\alpha }(1-f_{\text{esc}%
})q(H)\Phi (\nu _{\text{Ly}\alpha }-\nu ,z),
\end{equation}%
where $c_{\text{Ly}\alpha }$ represents the energy output per Ly$\alpha $
photon. The distribution of stellar masses, defined by the Initial Mass
Function (IMF), significantly impacts the emission. The Salpeter IMF ($\phi
(M)\propto M^{-2.35}$) and the Larson IMF \cite{C6,C7}, 
\begin{equation}
\phi (M)\propto M^{-1}\left( 1+\frac{M}{M_{c}}\right) ^{-1.35},
\end{equation}%
with $M_{c}$ as a characteristic mass scale, reflect differing biases toward
massive stars. These processes and equations highlight the critical
interplay between early star formation, photon scattering, and the evolving
IGM in shaping the CIB during the cosmic dark ages.

\begin{figure}[h]
\centering
\includegraphics[width=0.6\textwidth]{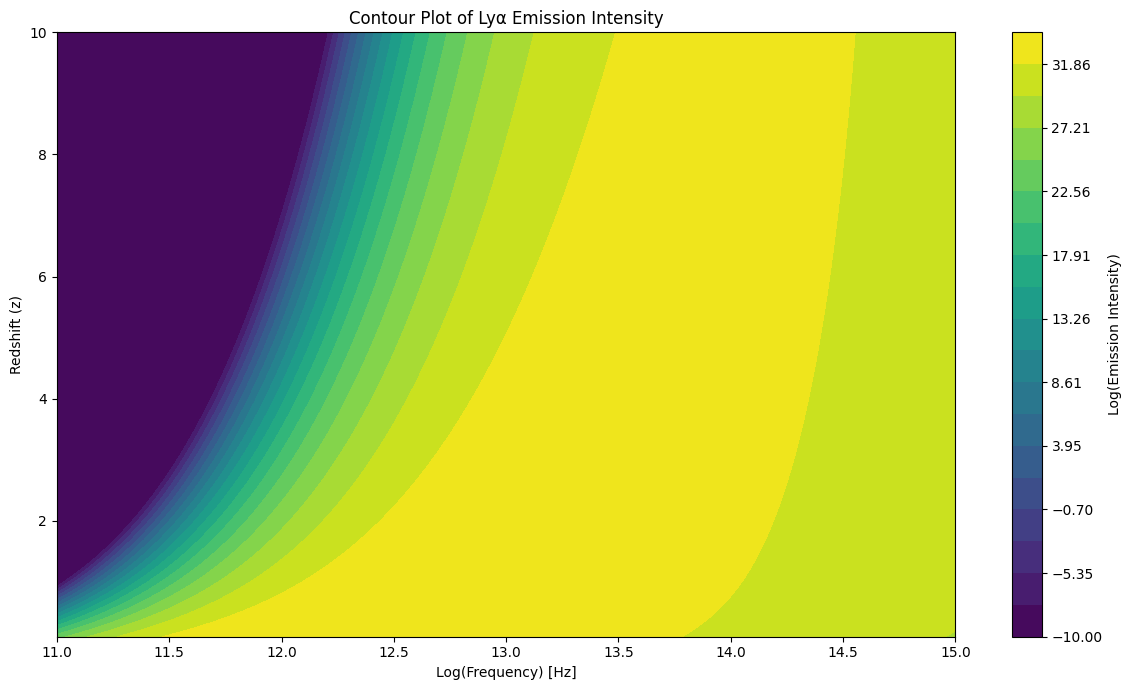} 
\caption{ Evolution of Ly$\protect\alpha $ Emission Line }
\end{figure}
Fig. (\ref{fig:0b}) illustrates the complex behavior of Ly$\alpha $ emission
intensity as a function of frequency and redshift, capturing how the
interplay between cosmic expansion and astrophysical processes shapes the
spectral energy distribution. The logarithmic frequency axis spans a wide
range, from $10^{11}Hz$ to $10^{15}Hz$, covering the typical scales where Ly$%
\alpha $ photons contribute, while the redshift extends from $z=0.1$,
representing the relatively recent universe, to $z=10$, corresponding to the
epoch of early galaxy and star formation. At high redshifts $(z\geq 5)$, the
intensity contours show a concentration at lower frequencies due to
significant redshifting of the emitted Ly$\alpha $ photons. The emission
intensity peaks at frequencies corresponding to the rest-frame Ly$\alpha $
transition redshifted by the factor $(1+z)^{-1}$, creating a strong diagonal
trend in the plot. This regime reflects the dominance of primordial ionizing
sources, such as early stars and galaxies, which emit photons that scatter
through the surrounding intergalactic medium. For Intermediate Redshifts $%
(2\leq z\leq 5)$, the emission intensity exhibits a broader spread across
frequencies, with a gradual weakening of the peak intensity. This broadening
is attributed to the combined effects of cosmic expansion, which reduces the
energy density of Ly$\alpha $ photons, and the partial reionization of the
IGM, which alters the scattering cross-section. The contours become less
sharp, indicating a smoother transition of intensity across frequencies.
Finally at Low Redshift $(z\leq 2)$, the contours shift further toward the
low-frequency regime, with a dramatic decline in peak intensity. This
reflects the universe's continued expansion, which increasingly dilutes the
energy of Ly$\alpha $ photons, as well as the reduced contribution of new
ionizing sources. The declining intensity also mirrors the reduced star
formation rates in the universe's late stages, contributing less to the Ly$%
\alpha $ photon population. The emission intensity drops sharply at high
frequencies ($\nu \geq 10^{14}Hz$) due to the exponential cutoff in the
scattering profile $\Phi (\nu ,z)$, governed by the redshift-dependent
characteristic frequency $\nu _{s}(z)$. In the mid-range ($10^{12}\leq \nu
\leq 10^{14}Hz$), intensity varies significantly with redshift, shaped by
the redshifted Ly$\alpha $ peak and the baryon density parameter $\Omega
_{b}h^{2}$. At low frequencies ($\nu \leq 10^{12}Hz$), intensity diminishes
uniformly across redshifts due to reduced scattering efficiency, reflecting
the spectral energy distribution of the ionizing sources. the results
captures the Ly$\alpha $ emission evolution as a probe of cosmological
history illustrating how early star formation epochs at high $z$ imprint
their signatures on the IGM, which are then redshifted and attenuated as the
universe ages. The drop in intensity at lower redshifts highlights the
diminishing role of new ionizing sources, making Ly$\alpha $ photons crucial
tracers of reionization and large-scale structure formation.

\section{ [C II] Emission As a Metallicity and star Formation Tracer}

\label{sec3} The [C II] $158\mu m$ line serves as a powerful tracer of both
metallicity and star formation in galaxies due to its origin in singly
ionized carbon $\left( C^{+}\right) $, a key component of the interstellar
medium. Its connection to metallicity arises from the fact that carbon is
produced in stars and enriched in the ISM through stellar winds and
supernovae, with its abundance increasing as galaxies evolve chemically. The
strength of the [C II] line depends on carbon abundance and cooling
efficiency, making it particularly bright in metal-poor environments where
alternative cooling mechanisms, such as CO, are less effective. Observations
have demonstrated that the [C II]-to-total infrared (TIR) luminosity ratio
decreases in higher metallicity regions due to dust reprocessing effects 
\cite{B1}. Additionally, the [C II]/CO ratio is often used to identify
low-metallicity systems where CO is under-abundant \cite{B2}. As a tracer of
star formation, the [C II] line is strongly linked to photodissociation
regions (PDRs), where UV photons from young, massive stars heat the gas,
with the subsequent emission reflecting the balance between UV heating and
cooling. Empirical studies have shown a strong correlation between [C II]
luminosity and star formation rate (SFR), particularly in moderate to high
star formation regimes \cite{B3}. However, this relationship can be
complicated in extremely dusty or metal-rich environments, where other
cooling mechanisms may dominate. The far-infrared nature of the [C II] line
makes it highly effective in penetrating dust, providing a clearer view of
star-forming regions compared to UV or optical tracers, which are often
obscured. Despite its advantages, the [C II] line can face limitations, such
as saturation in dusty systems or dominance by PDR contributions, requiring
additional spectral lines for contextual interpretation. Together, the [C
II] line is a critical tool for understanding the chemical evolution and
star formation activity of galaxies, particularly in the high-redshift
universe, where it remains bright and detectable even in dust-obscured
conditions \cite{B4}. 
\begin{figure}[h]
\centering
\includegraphics[width=0.8\textwidth]{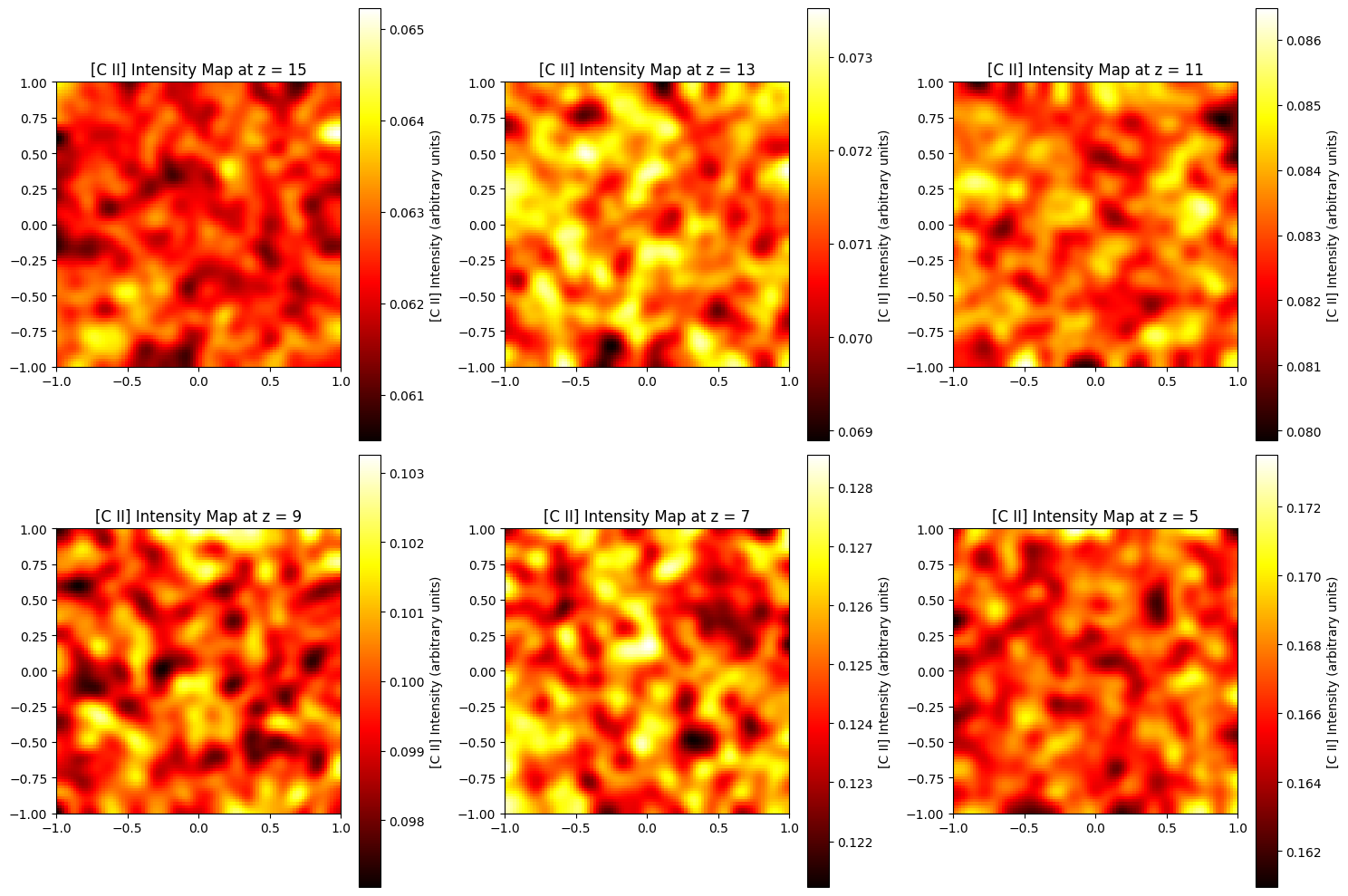} 
\caption{ [C II] Intensity across different redshifts }
\label{fig:1}
\end{figure}
The simulated [C II] intensity maps in Fig. (\ref{fig:1}) provide a valuable
perspective on the evolution of galaxies across redshifts, highlighting the
interconnected roles of star formation rate (SFR) and metallicity in shaping
observed emissions, the x-axis and y-axis represent spatial coordinates on a
2D grid. At higher redshifts $z\sim 15$, the maps show weaker intensities,
reflecting the early stages of galaxy formation when metallicities were
lower, and the ISM was dominated by primordial gas. As redshift decreases,
the simulated trend shows progressively enhanced [C II] intensities. This is
driven by rising star formation rates and the enrichment of the ISM through
stellar nucleosynthesis, both of which boost [C II] emission. These findings
align well with observational evidence that [C II] serves as a robust tracer
of star-forming regions, directly correlating with SFR and metallicity. In
the simulation, metallicity influences the abundance of carbon in the ISM,
which, upon ionization, produces the prominent [C II] 158 $\mu m$ emission
line. The maps also illustrate how cosmic evolution leads to brighter and
more widespread emission, consistent with the transition from the Epoch of
Reionization to the peak of star formation activity. These results are
directly comparable to real observations from ALMA and JWST, which have
already detected [C II] in galaxies at $z\sim 6$ and aim to push detections
to $z>10$. The use of Gaussian smoothing in the simulation accounts for
observational limitations, reflecting the spatial resolution challenges
faced in high-redshift studies. By connecting [C II] emission to SFR and
metallicity.

The [C II] $158\mu m$ line is a critical tool for tracing both metallicity
and star formation in galaxies, particularly in the context of high-redshift
studies. As it arises from singly ionized carbon $C^{+}$, its luminosity $L_{%
\left[ CII\right] }$\ is directly linked to the carbon abundance in the
interstellar medium, which scales with the galaxy's metallicity $Z$. This
relationship can be expressed as $L_{\left[ CII\right] }\propto C^{+}\propto
Z,$\ making it a reliable tracer of chemical evolution. Observationally, the 
$\left[ CII\right] \mathit{-to-CO}$ ratio, $R_{\left[ CII\right] /CO}=L_{%
\left[ CII\right] }/L_{CO}$, is often used to distinguish low-metallicity
environments, as CO formation is suppressed due to insufficient dust
shielding in these regions \cite{B2}. Additionally, the \textit{[C
II]-to-total infrared} ratio, $R_{\left[ CII\right] /TIR}=L_{\left[ CII%
\right] }/L_{TIR}$, provides insights into the cooling efficiency of carbon
relative to dust reprocessing, with $R_{\left[ CII\right] /TIR}$ decreasing
at higher metallicities where dust dominates \cite{B1}. \newline

The [C II] line is also a valuable tracer of star formation, as it primarily
originates in photodissociation regions (PDRs), where UV photons from young,
massive stars heat the surrounding gas. The [C II] luminosity scales
empirically with the star formation rate (SFR), often expressed as 
\begin{equation}
SFR\left( M_{\odot }yr^{-1}\right) =\alpha \left( \frac{L_{\left[ CII\right]
}}{10^{7}L_{\odot }}\right) ,
\end{equation}
where $\alpha $ is a calibration constant depending on the galaxy's dust
content and environment \cite{B3}. Furthermore, the balance between heating
and cooling in PDRs can be quantified through the photoelectric heating
efficiency $\epsilon $ is given by{},%
\begin{equation}
\epsilon =\frac{L_{\left[ CII\right] }+L_{\left[ OI\right] }}{L_{TIR}},
\end{equation}%
here\ $L_{\left[ OI\right] }$ represents the luminosity of the [O I]
fine-structure line. Higher $\epsilon $ values correspond to regions of
active star formation, with efficient UV heating and cooling \cite{B5}. In
high-redshift galaxies, the [C II] luminosity can be observed and corrected
for cosmological effects using the relationship%
\begin{equation}
L_{\left[ CII\right] ,obs}=\frac{L_{\left[ CII\right] ,int}}{4\pi
D_{L}^{2}\left( 1+z\right) },
\end{equation}%
where $D_{L}$\ is the luminosity distance and $z$ is the redshift. This
allows for the use of [C II] as a robust tracer of star formation and
metallicity in the early universe, even when traditional optical or UV
tracers are obscured by dust \cite{B4}. These mathematical frameworks
highlight the dual utility of the [C II] line in probing both the chemical
enrichment and the star formation activity of galaxies, especially under the
challenging observational conditions of the distant universe.

\begin{figure}[h]
\centering
\includegraphics[width=0.5\textwidth]{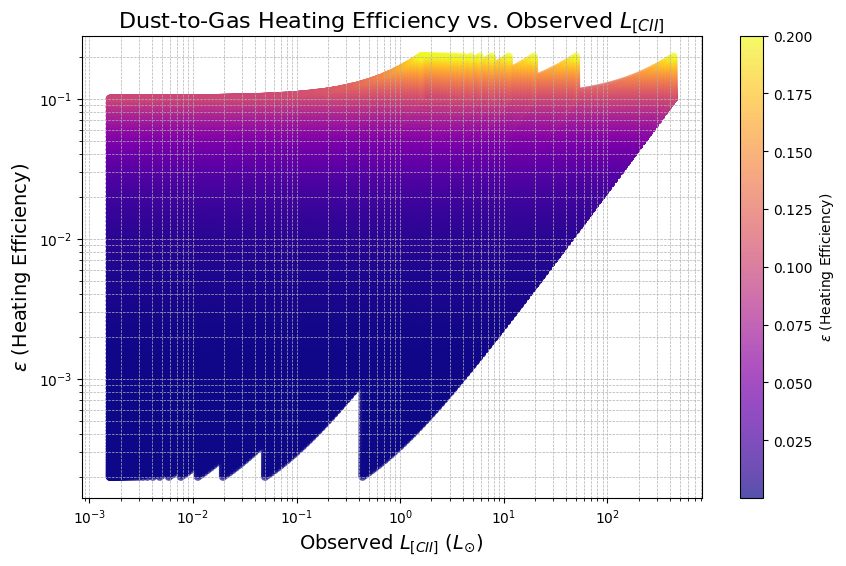} 
\caption{Dust-to-Gas Heating Efficiency of the [C II] $158\protect\mu m$
line }
\label{fig:2}
\end{figure}

Fig.(\ref{fig:2}) provides a detailed simulation of galaxy properties to
explore the relationship between dust-to-gas heating efficiency and observed
[C II] luminosity. Using intrinsic luminosities of [C II] and [OI] lines and
their dependence on the total infrared luminosity $L_{TIR}$. The model
incorporates cosmological effects by calculating the luminosity distance $%
D_{L}$ based on the redshift and assumes a flat universe. The dust-to-gas
heating efficiency relates the energy emitted in cooling lines ([C II] and
[O I]) to the total infrared energy emitted by dust. Observed [C II]
luminosity is derived by accounting for the term $1/D_{L}^{2}\left(
1+z\right) $, providing insights into how intrinsic properties of galaxies
are altered by their distance and the expansion of the universe.

The visualization reveal the dependency of heating efficiency on cooling
line emissions and their detectability. This approach mirrors observational
strategies used in studies of high-redshift galaxies, such as with ALMA or
JWST, to infer SFRs, ISM conditions, and dust properties. The color gradient
indicates variations in heating efficiency, with brighter regions (yellow)
representing higher efficiencies and darker regions (blue) representing
lower efficiencies. The observed trend shows that $\epsilon $ tends to
increase with increasing $L_{[CII]}$, but this relationship is not linear
and exhibits significant scatter at low luminosities. This suggests that
higher [CII] luminosities generally correlate with more efficient gas
heating, which could reflect variations in environmental conditions, such as
the density, temperature, or metallicity of the ISM. The irregularities and
scatter at low luminosities may indicate complex, non-linear processes
influencing the heating efficiency, possibly linked to localized star
formation or the interplay of dust and gas in different regions. 
\begin{figure}[h]
\centering
\includegraphics[width=0.7\textwidth]{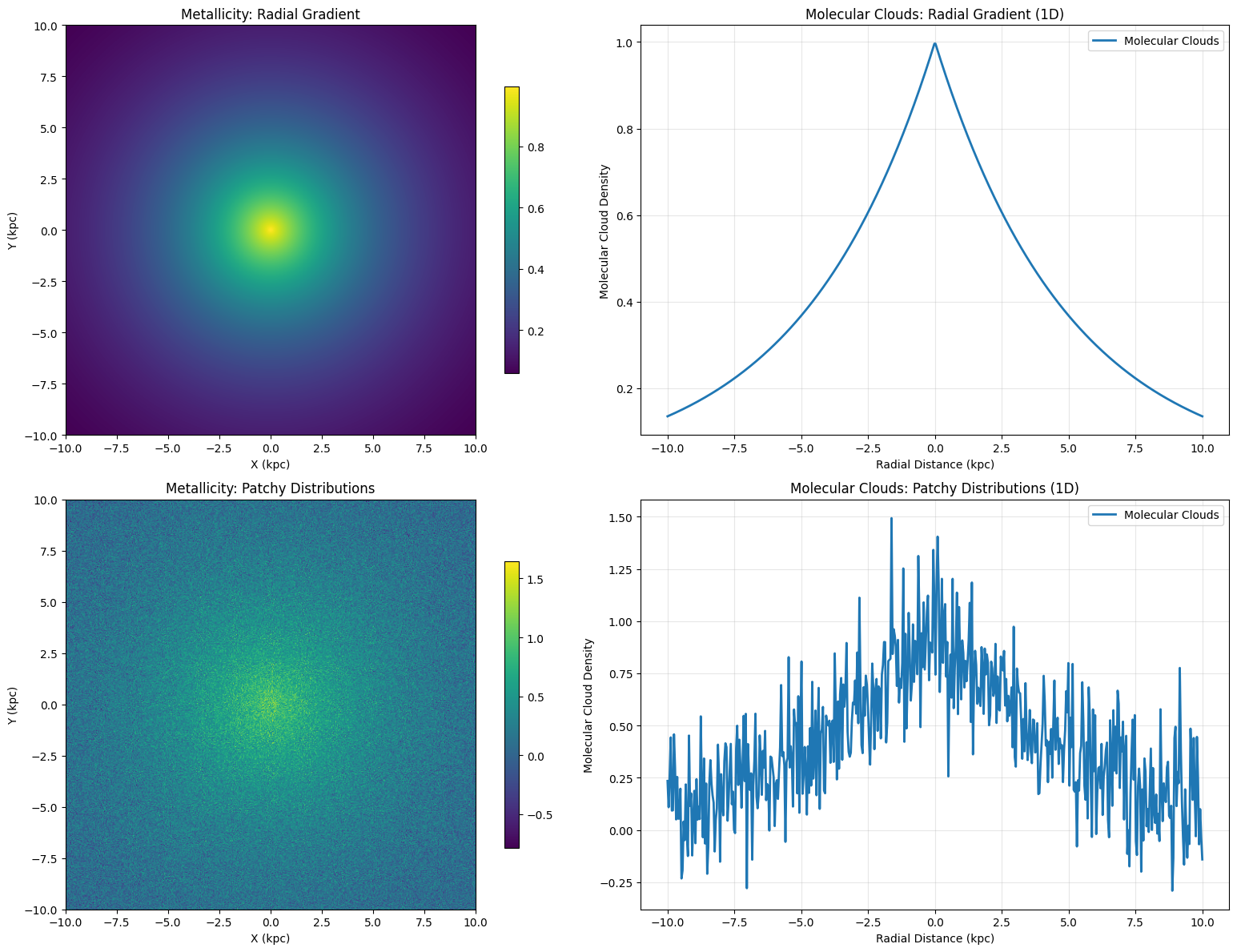} 
\caption{ Metallicity and Molecular Cloud Distributions: Radial Gradient vs.
Patchy Models }
\label{fig:3}
\end{figure}

Fig.(\ref{fig:3}) examines the spatial interplay between metallicity and
molecular cloud distributions in galaxies through two models: the Radial
Gradient and Patchy Distributions. The radial gradient model assumes an
exponential decline in metallicity and molecular cloud density with radius,
cloud, reflecting the enrichment of the ISM in central regions due to
sustained star formation and efficient recycling of stellar materials. This
model captures large-scale, axisymmetric trends, with inner regions hosting
denser molecular clouds and higher metallicity, resulting in stronger [CII]
emission line driven by efficient cooling and UV heating in
photodissociation regions (PDRs). Conversely, the patchy distribution model
introduces local stochastic variations, mimicking turbulent feedback effects
such as supernova explosions and stellar winds that disrupt the ISM,
creating fragmented molecular clouds and localized metallicity enhancements.
This model explains the clumpy nature of [CII] emission in high-feedback
environments, such as starburst regions or interacting galaxies, where
molecular clouds are dynamically formed and destroyed. Together, these
models describe both the global ISM behavior and localized feedback-driven
variability, offering a robust framework for interpreting the spatial
distribution of [CII] line emission and its connection to star formation.

\section{Comparative Analysis of Ly$\protect\alpha $ and [C II] in
High-Redshift Galaxies}

\label{sec4} Understanding the formation and evolution of high-redshift
galaxies during the epoch of reionization relies heavily on emission lines
such as Lyman-alpha and [C II]. These lines serve as critical diagnostics of
star formation, interstellar medium conditions, and feedback processes.Ly$%
\alpha $ originating from hydrogen recombination, provides insights into the
ionization state of the intergalactic medium \cite{E1}, while [C II], a
prominent cooling line in photo-dissociation regions (PDRs), acts as a
tracer of star formation and metallicity. Here, we present a theoretical
framework and a computational approach to simulate these emission lines,
bridging astrophysical theory and observational predictions. The observed
flux $F_{Ly\alpha }$ of the Ly$\alpha $ emission line is related to the
intrinsic luminosity $L_{Ly\alpha }$ by,%
\begin{equation}
F_{Ly\alpha }=\frac{L_{Ly\alpha }}{4\pi d_{L}^{2}},
\end{equation}%
where $d_{L}$ is the luminosity distance at redshift $z.$\ In a flat $%
\Lambda -CDM$\ cosmology\ is given by $d_{L}=\left( 1+z\right) \cdot \frac{c%
}{H_{0}}\int_{0}^{z}\frac{dz^{\prime }}{\sqrt{\Omega _{m}\left( 1+z^{\prime
}\right) ^{3}+\Omega _{\Lambda }}}$,\ with $c$\ as the speed of light $H_{0}$%
\ as the Hubble constant, $\Omega _{m}$ and $\Omega _{\Lambda }$\ as the
matter and dark energy density parameters. Observations and models suggest
that $L_{Ly\alpha }\propto SFR$\ as Ly$\alpha $ emission scales with the
ionizing photon output from young, massive stars \cite{X3, X4}. The [C II]
emission line at a rest-frame frequency is a major cooling line for the ISM,
especially in PDRs illuminated by young stars. The observed flux $F_{[CII]}$%
, depends on the intrinsic luminosity $L_{[CII]}$ and is similarly given by,%
\begin{equation}
F_{[CII]}=\frac{L_{[CII]}}{4\pi d_{L}^{2}}.
\end{equation}%
Empirical studies, such as \cite{X5}, show a correlation between $L_{[CII]}$
and SFR, indicating that $L_{[CII]}\propto SFR$. For both emission lines,
the rest-frame wavelength or frequency shifts into the observed frame due to
cosmological redshift,%
\begin{eqnarray}
\lambda _{obs} &=&\lambda _{rest}\cdot \left( 1+z\right) ,  \notag \\
\nu _{obs} &=&\frac{\nu _{rest}}{1+z}.
\end{eqnarray}%
The $Ly\alpha $ falls in the near-infrared regime, while [C II] appears in
the millimeter regime for high-redshift galaxies.

This study models the Ly$\alpha$ flux to decline with redshift due to
increased IGM opacity, described by an attenuation factor $\exp (-\tau
_{IGM})$. For [C II], the emission depends on SFR and ISM properties, with
metallicity influencing the cooling efficiency.

This theoretical framework provides a path for understanding the connections
between galaxy properties (SFR, metallicity, redshift) and emission line
fluxes during the epoch of reionization. By focusing on intrinsic
astrophysical relationships, the analysis serves as a baseline for exploring
galaxy evolution and ISM conditions especially for testing hypotheses about
early galaxy formation. 
\begin{figure}[h]
\centering
\includegraphics[width=0.99\textwidth]{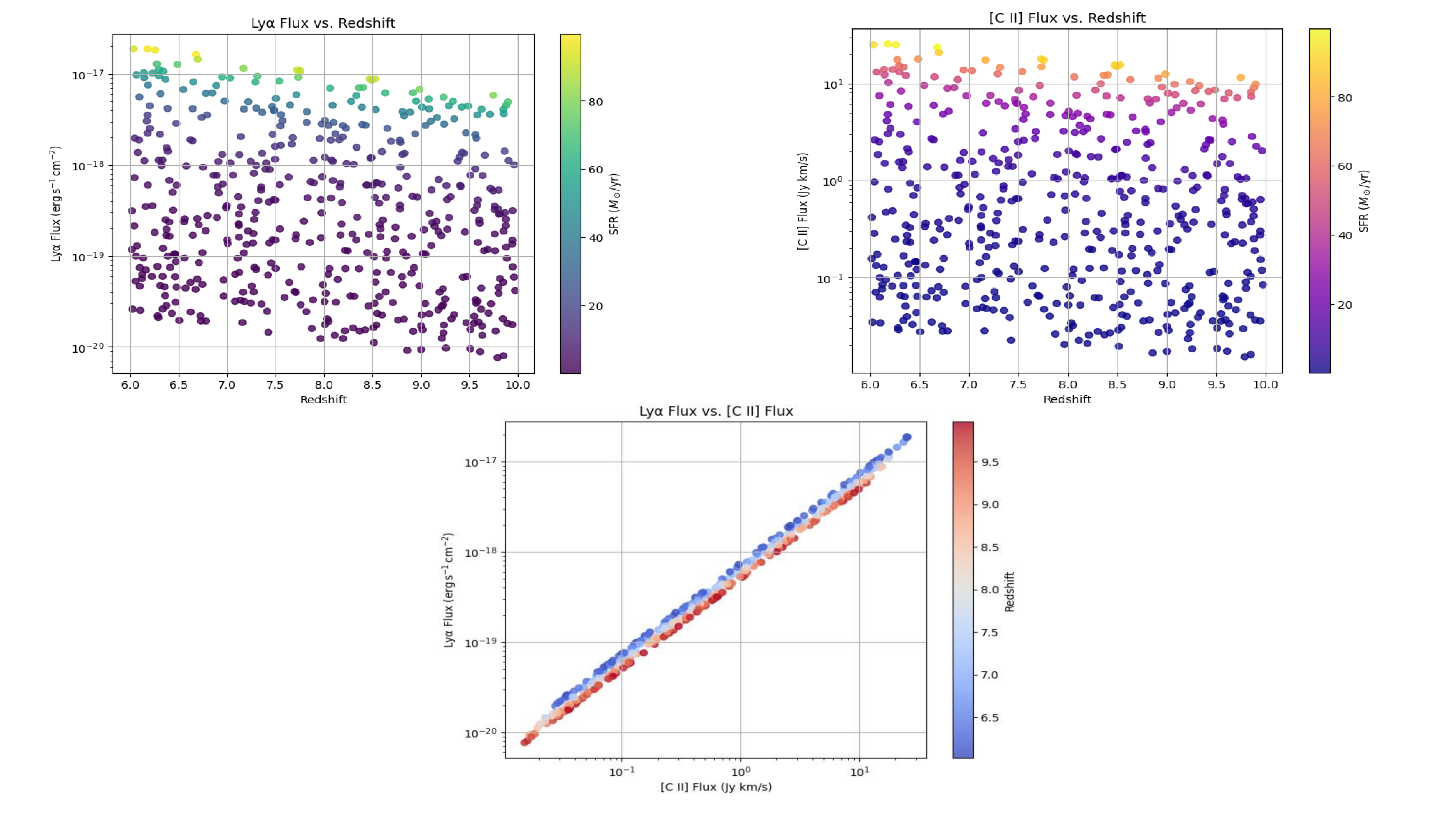} 
\caption{ Metallicity and Molecular Cloud Distributions: Radial Gradient vs.
Patchy Models }
\label{fig:4}
\end{figure}
The provided results in Fig.(\ref{fig:4}) visualizes through scatter plots
the relationship between fluxes, redshifts, and SFRs to simulate the
theoretical emission properties of high-redshift galaxies during the epoch
of reionization, focusing on the Ly$\alpha $ and [C II] lines. Using a flat $%
\Lambda $-CDM cosmology, the code predicts the fluxes of these lines for
galaxies with redshifts ranging from $z\sim 6$ to $z\sim 10$ and star
formation rates (SFRs) spanning $0.1$ to $100M_{\odot }yr^{-1}$. The $%
Ly\alpha $ flux is modeled as inversely proportional to the square of the
luminosity distance, with attenuation due to the intergalactic medium (IGM).
The [C II] flux is similarly scaled, incorporating its dependence on SFR and
redshift, while assuming a simplified metallicity factor. The flux$\ $%
decreases with increasing redshift, primarily due to the growing optical
depth of the IGM, which attenuates the emission. Conversely, the [C II] flux
remains more stable, as it is less affected by the IGM and directly tied to
the SFR and ISM conditions. Moreover, the $Ly\alpha $ line, is redshifted
into the near-infrared, while the [C II] line shifts into the millimeter
range. These shifts follow the cosmological redshift relations and enable
the placement of these lines in observational spectra for high-redshift
galaxies. The [C II] flux calculation assumes a uniform metallicity,
emphasizing its theoretical dependence on SFR alone, the results helps
understanding of how $Ly\alpha $ and [C II] emission lines trace galaxy
properties in the early universe, serving as a highlight the distinct roles
of these lines in probing galaxy evolution during cosmic history.

\section{Conclusion}

\label{sec5} This paper aims to explore the critical processes and
observational signatures of the Cosmic Dark Ages, focusing on the transition
from a neutral, dark universe to one illuminated by the first stars and
galaxies. By leveraging theoretical models and simulations, the study
examines how key emission lines such as Ly$\alpha $ and [C II] 158 $\mu m$
trace star formation, metallicity, and intergalactic medium conditions,
providing insights into the epoch of reionization. Firstly, we examined the
spatial and emission profiles reveal the dominance of high-mass,
low-metallicity stars in driving nebular emissions and radiative output
during the early stages of galaxy formation. The simulated spatial
distributions underscore the rarity of massive stars but highlight their
critical role in reionization and enriching the IGM. On the other hand, the
evolution of Ly$\alpha $ emission intensity across redshifts and frequencies
captures the interplay between cosmic expansion and photon scattering in the
IGM. At high redshifts, the sharp contours of Ly$\alpha$ emission reflect
the influence of primordial ionizing sources, while at lower redshifts, the
emission weakens due to declining star formation rates and IGM transparency.
Moreover, the [C II] intensity maps and their connection to star formation
and metallicity emphasize the line's effectiveness in probing galaxy
properties. The results demonstrate the evolution of [C II] emission from
weak, localized regions in low-metallicity environments to brighter,
widespread emissions as galaxies evolve chemically. Furthermore, the
relationship between [C II] luminosity, star formation rates, and
dust-to-gas heating efficiency provides a comprehensive view of the ISM's
radiative cooling processes. These results align with observational trends,
reinforcing the role of [C II] as a reliable tracer of early star-forming
galaxies. Then, further invertigations shows that the radial gradient and
patchy distribution models highlight the dual influence of large-scale ISM
enrichment and localized feedback-driven variability on [C II] emissions.
These models capture the balance between steady enrichment processes and
stochastic feedback effects, providing a framework for interpreting emission
line clumpiness in high-redshift galaxies. Finally, the contrasting
behaviors of Ly$\alpha$ and [C II] fluxes across redshifts reveal their
complementary roles in tracing early galaxy properties. While Ly$\alpha$ is
highly sensitive to IGM opacity and declines sharply with redshift, [C II]
remains stable and directly correlates with star formation and metallicity.
Overall, this study bridges theoretical predictions with simulation,
offering a robust framework for interpreting high-redshift galaxy evolution.
The findings provide valuable guidance for upcoming observations enhancing
our understanding of the universe's transition from the Cosmic Dark Ages to
the era of reionization.

\section{Aknowledgments}

G. Otalora acknowledges Direcci\'on de Investigaci\'on, Post-grado y
Transferencia Tecnol\'ogica de la Universidad de Tarapac\'a for financial
support through Proyecto UTA Mayor 4737-24.


\end{document}